\documentclass[aps,pra,amsmath,twocolumn]{revtex4}

\usepackage{amsmath,amssymb}
\usepackage{graphicx}
\usepackage{color}
\usepackage{xspace}
\usepackage{mathtools}

\newcommand{\bra}[1]{\langle #1|}
\newcommand{\ket}[1]{|#1\rangle}
\newcommand{\braket}[2]{\langle #1 | #2 \rangle}

\newcommand{\pder}[2]{\frac{\partial #1}{\partial #2}}
\newcommand{\der}[2]{\frac{d #1}{d #2}}

\newcommand{\QAO}{\textsc{qao}\xspace}

\renewcommand{\epsilon}{\varepsilon}

\preprint{Preprint}

\begin{document}

\title{Evolution-Time Dependence in Near-Adiabatic Quantum Evolutions}

\author{Lucas~T.~Brady}
\affiliation{{Department of Physics, University of California, Santa Barbara, CA 93106-5110, USA}}
\author{Wim~van~Dam}
\affiliation{{Department of Computer Science, Department of Physics, University of California, Santa Barbara, CA 93106-5110, USA}}

\date{\today}
%%%%%%%%%%%%%%%%%%%%%%%%%%%%%%%%%% Abstract
\begin{abstract}
We expand upon the standard quantum adiabatic theorem, examining the time-dependence of quantum evolution in the near-adiabatic limit.  We examine a Hamiltonian that evolves along some fixed trajectory from $\hat{H}_0$ to $\hat{H}_1$ in a total evolution-time $\tau$, and our goal is to determine how the final state of the system depends on $\tau$.  If the system is initialized in a non-degenerate ground state, the adiabatic theorem says that in the limit of large $\tau$, the system will stay in the ground state.  We examine the near-adiabatic limit where the system evolves slowly enough that most but not all of the final state is in the ground state, and we find that the probability of leaving the ground state oscillates in $\tau$ with a frequency determined by the integral of the spectral gap along the trajectory of the Hamiltonian, so long as the gap is big.  If the gap becomes exceedingly small, the final probability is the sum of oscillatory behavior determined by the integrals of the gap before and after the small gap.  We confirm these analytic predictions with numerical evidence from barrier tunneling problems in the context of quantum adiabatic optimization.
\end{abstract}

\maketitle

%%%%%%%%%%%%%%%%%%%%%%%%%%%%%%%%%% Introduction
\section{Introduction}
\label{sec:intro}

The Quantum Adiabatic Theorem \cite{BornFock} is a powerful tool for analyzing dynamical quantum systems.  For slowly evolving Hamiltonians it ensures the system will closely track its originally initialized energy state throughout the entire evolution.  The key point of the adiabatic theorem is that it also gives a condition for how slowly the system needs to evolve.

The adiabatic theorem is notable for many applications in physics and chemistry.  Of specific note for our purposes are Quantum Adiabatic Optimization \cite{Farhi2000} and Quantum Annealing which are a class of quantum algorithms for solving optimization problems.  These algorithms initialize the system in an easily prepared eigenstate and rely on the adiabatic theorem to evolve it into a desired state under the influence of a suitably designed Hamiltonian.  Quantum adiabatic computing is universal for quantum computing in general \cite{Aharonov}, but it is currently a matter of debate \cite{Farhi2002, Farhi2008, Martonak, Hastings, Heim, Boxio2, Battaglia, Crosson, Harrow, Muthukrishnan, Brady, Kong, Jiang} how much of that power can be captured by the model used in many applications, such as the D-Wave machine \cite{DWave}.

In this study, we will focus on the near-adiabatic regime of quantum evolution and quantum annealing, where the system evolves slowly enough to mostly stay in the desired eigenstate but with noticeable leakage.  Non-adiabatic evolution has garnered much interest because it can potentially lead to speed-ups that the adiabatic theorem does not account for.  In quantum annealing, recent work has focused on a rapid diabatic speed-up in certain barrier tunneling models \cite{Kong, Muthukrishnan, Brady3}.

While we couch our findings in the language of quantum computing, our results are more general.  Many of our assumptions and approximations are general and can apply to a large number of problems and settings.  Quantum Adiabatic Optimization relies on ground state evolution, and even this condition can be relaxed in our results.

In the near-adiabatic limit, our findings show that in the absence of a small spectral gap, the probability of transitioning out of the initialized state oscillates as a function of the total evolution time, $\tau$.  Furthermore, we confirm previous results \cite{Grandi,Lidar,Wiebe} that show that the frequency of this oscillation depends on the integral of the spectral gap over the evolution.

Our new results add a layer of depth by considering the case where the spectral gap becomes small during one portion of the evolution, as would occur in a Landau-Zener avoided crossing \cite{Landau, Zener}.  This crossing is localized, so that it only effects the system during a short period of time.  We find that the avoided crossing effectively splits the evolution in two, resulting in a superposition of oscillatory behavior in $\tau$.

This article is structured with a basic overview of the Quantum Adiabatic Theorem and its context in quantum computing in Section \ref{sec:adi_theorem}.  In Section \ref{sec:setup}, we develop much of the mathematical machinery that will be used in later sections.

We study the large gap regime in Section \ref{sec:large_gap}, examining how the oscillatory transition probability behavior arises, and we back up these results with numerical evidence from quantum algorithm settings.  In Section \ref{sec:small_gap}, we add in an avoided level crossing and explore analytically and numerically how this leads to a more complicated superposition of oscillatory behavior in total evolution time, $\tau$.  Section \ref{sec:grover} shows an application of the large gap oscillations to the adiabatic version of Grover's search.  Finally we review our conclusions and discuss future avenues of study in Section \ref{sec:conc}.

\section{Adiabatic Theorem}
\label{sec:adi_theorem}

Suppose we have some quantum system obeying the Schr\"odinger equation
\begin{equation}
      i\der{}{t}\ket{\psi} = \hat{H}(t)\ket{\psi},
\end{equation}
where time runs between $t=0$ and $\tau$, the evolution time of the system or the run time in a quantum computing setting.  The Hamiltonian should follow the same trajectory even for different values of $\tau$, so we can use the ``normalized time'' $s\in[0,1]$ to determine where in the Hamiltonian's evolution we are.

The normalized time $s$ relates to the actual time through $t=s\tau$.  Since the form of the Hamiltonian depends only on $s$ and not $\tau$, we can rewrite the Schr\"odinger equation as
\begin{equation}
      \label{eq:s_Schro}
      i\der{}{s}\ket{\psi} = \tau\hat{H}(s)\ket{\psi}.
\end{equation}
Now all the evolution time information has been pulled out into one parameter, $\tau$, which we can vary to run the evolution more slowly or quickly.

\subsection{Quantum Adiabatic Theorem}

The Quantum Adiabatic Theorem is an old result first attributed to Born and Fock \cite{BornFock} but not treated fully rigorously until more recently (e.g. \cite{Jansen, Lidar}).  The theorem concerns systems with a time dependent Hamiltonian, $\hat{H}(s)$, that is initialized in the $i$th eigenstate.  If the $i$th eigenstate has a non-zero spectral gap, $\Delta(s)$, separating it from other eigenstates for the entire time evolution, then a sufficiently slow evolution of the Hamiltonian will keep the system in the $i$th eigenstate.

The key idea in the adiabatic theorem is how slowly the system must be evolved.  Specifically, how large must $\tau$ be to ensure that a significant portion of the probability remains in $i$th eigenstate.  An oft quoted folklore result is that the adiabatic theorem holds if
\begin{equation}
      \label{eq:adi_condition}
      \tau \gg \int_0^1 ds\,\frac{|\bra{\varphi_0}\der{\hat{H}}{s}\ket{\varphi_1}|}{\Delta(s)^2},
\end{equation}
where $\ket{\varphi_i}$ is the $i$th energy state.

This adiabatic condition is not the full rigorous condition for adiabatic evolution \cite{Jansen, Lidar, Reichardt}, but it is sufficient for most cases, especially in quantum computing.  Additionally, the more rigorous versions of the adiabatic condition still depend polynomially on the inverse of the spectral gap, $\Delta(s)^{-1}$, and matrix norms of the Hamiltonian and its $s$ derivatives.

For our purposes, we will always focus on the ground state.  Therefore, the spectral gap $\Delta(s)$ will just be the energy difference between the first excited state and the ground state.

\subsection{Quantum Adiabatic Optimization}
\label{ssec:QAO}

One major application of the quantum adiabatic theorem is to quantum computing.  Quantum Adiabatic Optimization (\QAO) is a quantum algorithm introduced in \cite{Farhi2000} building upon previous quantum annealing models \cite{Finnila, Kadowaki}.  In \QAO, a quantum system is initialized in the ground state of a simple Hamiltonian, and the Hamiltonian is then adiabatically evolved into one with a ground state that solves a desired computational problem.  By measuring the final ground state, a solution to the computational problem can be obtained.

The original framing of \QAO \cite{Farhi2000} works with an initial Hamiltonian that is a sum of $\sigma_x$ terms on $n$ qubits
\begin{equation}
      \hat{H}_0 = \frac{1}{2}\sum_{i=1}^n \sigma_x^{(i)}
\end{equation}
and a final Hamiltonian which is diagonal in the computational, $z$, basis and depends on some cost function $f(z)$
\begin{equation}
      \hat{H}_1 = \sum_{z\in\{0,1\}^n} f(z)\ket{z}\bra{z}.
\end{equation}
The goal is to find a bit string that minimizes $f(z)$; therefore, we are looking for the ground state of $\hat{H}_1$.
The algorithm linearly interpolates between the two Hamiltonians in total time $\tau$:
\begin{equation}
      \hat{H}(s) = \left(1-s\right)\hat{H}_0+s\hat{H}_1
\end{equation}

\QAO relies on the adiabatic theorem to keep the system in the ground state, but quantum annealing in general can run this algorithm faster than the adiabatic theorem recommends.  This paper can be interpreted in terms of quantum annealing as describing how a non-adiabatic evolution effects the final success probability of the algorithm.  We examine this success probability as a function of $\tau$, and we find that the success probability depends greatly on the spectral gap.

In particular, we look at situations where the spectral gap remains large, except possibly in isolated regions where avoided-level crossings are allowed.  To this end, we study two different symmetric qubit problems.

\subsubsection{$n$-Qubit Barrier Tunneling Model}

The first model we examine is one that has been studied in numerous articles \cite{Farhi2000, Brady, Crosson, Harrow, Muthukrishnan, Jiang, Kong, Brady3, Reichardt, Brady2} with a final Hamiltonian cost function
\begin{equation}
      \label{eq:cost_function}
      f(z) = \mu|z| +b(|z|),
\end{equation}
where $|z|$ is the Hamming weight of the bit string $z$.  The function $b(|z|)$ is some localized barrier function that has width and height that scale with $n^{\alpha}$ and $n^{\beta}$, respectively, for some constants $\alpha$ and $\beta$.  Depending on the values of $\alpha$ and $\beta$, \QAO can adiabatically require run times, $\tau$, that grow with $n$ in constant, polynomial, or exponential ways \cite{Brady2, Harrow, Jiang}.  In this article, we take the barrier to be localized around $|z|=\frac{n}{4}$ and take the shape of the barrier to be binomial; though, neither of these choices are particularly relevant.

The problem determined by Eq.~\ref{eq:cost_function} has a final ground state at $|z|=0$, but in order to reach the final state, the instantaneous state must pass through or over the barrier given by $b(|z|)$.  Adiabatically this is accomplished by a tunneling event, and it appears in the spectrum as an avoided level crossing between the ground state energy and the $n$-fold degenerate first excited state.  Thus, this problem, exhibits a large spectral gap for most time that is well approximated by the $b(|z|)=0$ case, except in the vicinity of the barrier where the problem takes on the form of a Landau-Zener avoided crossing.  This problem is often studied in order to extract how much tunneling effects \QAO.

Since the Hamiltonian, $\hat{H}(s)$ is symmetric between qubits, the symmetric subspace fully describes the eigenspectrum of the system.  Therefore, this $2^n$ dimensional Hamiltonian can be simulated using an $n+1$ dimensional system, described by a tridiagonal matrix.  This reduction of the size of the system allows for efficient calculation of the spectrum and other properties of the system numerically, allowing much larger $n$ to be studied.

We utilize this simple barrier tunneling model in \QAO as an example of how our near-adiabatic evolution depends on the evolution or run time, $\tau$.  Our analytic approximations and results are general and independent of this specific computational problem, but we use it as a numerical example to verify our analytic results.

\subsubsection{Cubic Potential Model}

The second model we consider works with a final potential that is cubic but without an explicit barrier.  This model is the $p=3$ case of the $p$-spin model that has been used by numerous groups \cite{Jorg, Seki, Bapst, Seoane, Susa, Susa2}, and in the language of Hamming weight it is given by the final potential
\begin{equation}
      \label{eq:cubic_cost}
      f(z) = n\left(2\frac{|z|}{n}-1\right)^3.
\end{equation}

This cost function still has the all-zero bit string as its ground state, so the annealing evolution should still take the ground state from being localized around $|z|=n/2$ at $s=0$ to $|z|=0$ at $s=1$.
This problem does not include a barrier in the final potential but can still be visualized as a barrier tunneling problem in a semi-classical large-$n$ limit, using such methods as the Villain transformation\cite{Brady2}.  Notably, J\"org \emph{et al.}~\cite{Jorg} showed that if the exponent $p\geq 3$, then the spectral gap becomes exponentially small in this problem.  Therefore, finding the ground state of this cost function through quantum adiabatic optimization is a difficult task.

Many of the useful properties from the $n$-qubit barrier model also carry over to this system.  The final cost function remains symmetric between qubits, so symmetry simplifications can be employed to make this problem numerically tractable to solve for large $n$.

\section{Setup}
\label{sec:setup}

We will start with the normalized time, $s$, version of the Schr\"odinger equation, Eq.~\ref{eq:s_Schro}.  The next step involves rewriting the equation in the eigenbasis of the Hamiltonian.  We take the instantaneous eigenbasis to be given by $\ket{\varphi_j(s)}$ with associated eigenenergies $\lambda_j(s)$.  Then a general state of our system is written as
\begin{equation}
      \ket{\psi(s,\tau)} = \sum_{j} C_j(s,\tau) \ket{\varphi_j(s)}.
\end{equation}

In terms of the eigenbasis the Schr\"odinger equation gives
\begin{equation}
      \label{eq:before_approx}
      i\left[\der{C_k}{s}+\sum_{j}C_j \bra{\varphi_k}\der{}{s}\ket{\varphi_j}\right] = \tau \lambda_k C_k.
\end{equation}
We want to know how the system evolves if we start the system in the ground state at $s=0$.  These same arguments work for higher excited states as well, but for our purposes, the ground state is sufficient and simpler.  The adiabatic theorem tells us that if $\tau$ is large enough, we remain in the ground state.  We relax the adiabatic condition slightly and allow near-adiabatic evolution where the majority of the state remains in $\ket{\varphi_0(s)}$, but some small amount leaks into the first excited states, with all other states being essentially unvisited.

We require the ground state to be nondegenerate, but the first excited eigenstate can be degenerate.  We denote this possible $m$-fold degeneracy with a superscript $C_{1}^{(a)}$.

We want to restrict down Eq.~\ref{eq:before_approx} to just those probability amplitudes that are assumed to be relevant, namely, those close to the ground state.  In doing this, we can remember that $\bra{\varphi_j}\der{}{s}\ket{\varphi_j}=0$.  We also shift the Hamiltonian by an overall ($s$-dependent) constant so that $\lambda_0=0$, which means that $\lambda_{1} = \Delta$ is just given by the spectral gap.  Using all this information, we obtain the following coupled equations for the relevant amplitudes
\begin{equation}
      i\der{C_0}{s} + i\sum_{a=1}^m C_{1}^{(a)}\bra{\varphi_0}\der{}{s}\ket{\varphi_{1}^{(a)}} = 0
\end{equation}
\begin{align}
      i\der{C_1^{(a)}}{s} &+ iC_0\bra{\varphi_1^{(a)}}\der{}{s}\ket{\varphi_0}\nonumber\\
      &+i\sum_{b\neq a} C_1^{(b)}\bra{\varphi_1^{(a)}}\der{}{s}\ket{\varphi_1^{(b)}} = \tau\Delta C_1^{(a)}
\end{align}

For $\bra{\varphi_1^{(a)}}\der{}{s}\ket{\varphi_1^{(b)}}$, we can freely choose our basis within the degenerate eigenspace, and it is possible and desireable to choose our degenerate eigenbasis such that $\bra{\varphi_1^{(a)}}\der{}{s}\ket{\varphi_1^{(b)}} = 0$ for all $a$ and $b$.  

Just from the definition of the eigenvalues of $\hat{H}(s)$ and the orthonormality of its eigenvectors, we can look at 
\begin{align}
      \pder{}{s}\left(\hat{H}\ket{\varphi_i}\right) = \pder{}{s}\left(\lambda_i \ket{\varphi_i}\right)\\
      \dot{\hat{H}}\ket{\varphi_i} + \hat{H}\pder{}{s}\ket{\varphi_i} = \pder{\lambda_i}{s}\ket{\varphi_i}+\lambda_i\pder{}{s}\ket{\varphi_i}\nonumber.
\end{align}
We can look at the inner product of this time derivative with an eigenstate $j\neq i$ and utilize the orthonormatility of the eigenstates
\begin{align}
      \bra{\varphi_j}\dot{\hat{H}}\ket{\varphi_i} + \bra{\varphi_j}\hat{H}\pder{}{s}\ket{\varphi_i} &= \pder{\lambda_i}{s}\braket{\varphi_j}{\varphi_i}+\lambda_i\bra{\varphi_j}\pder{}{s}\ket{\varphi_i}\nonumber\\
      \bra{\varphi_j}\dot{\hat{H}}\ket{\varphi_i} &= (\lambda_i-\lambda_j)\bra{\varphi_j}\pder{}{s}\ket{\varphi_i}.
\end{align}
which means that
\begin{align}
      &\bra{\varphi_0(s)}\der{}{s}\ket{\varphi_1^{(a)}(s)} = -\bra{\varphi_1^{(a)}(s)}\der{}{s}\ket{\varphi_0(s)} \\
      = &\frac{\bra{\varphi_0(s)}\der{\hat{H}}{s}\ket{\varphi_1^{(a)}(s)}}{\Delta(s)} \equiv \frac{\gamma_a(s)}{\Delta(s)}.
\end{align}

Thus, our differential equations can be reduced to
\begin{align}
      \label{eq:C0eq}
      \der{C_0}{s} + \sum_{a=1}^m C_{1}^{(a)}\frac{\gamma_a}{\Delta} &= 0\\
      \label{eq:C1eq1}
      i\der{C_1^{(a)}}{s} - iC_0 \frac{\gamma_a}{\Delta} &= \tau\Delta C_1^{(a)}.
\end{align}

At this point it is obvious that all the $m$-fold degenerate first excited states will behave in the exact same way.  Therefore, for notational convenience, we will drop the $a$ index and make the $m$-fold degeneracy explicit in these differential equations:

\begin{align}
      \label{eq:C0eq}
      \der{C_0}{s} + m C_{1}\frac{\gamma}{\Delta} &= 0\\
      \label{eq:C1eq1}
      i\der{C_1}{s} - iC_0 \frac{\gamma}{\Delta} &= \tau\Delta C_1.
\end{align}

\section{Large Gap}
\label{sec:large_gap}

\subsection{Analytic Approximation}
\label{ssec:large_anal}

Now, we take the near-adiabatic limit by assuming that the vast majority of of the state remains in the ground state.  In this limit, we assume that $C_0\gg C_1$ so that Eq.~\ref{eq:C0eq} reduces to $\der{C_0}{s}=0$.  This reduction assumes that the gap $\Delta$ does not become too small.  If the gap becomes small, then this problem can be approximated using different techniques, such as the Landau-Zener transition in the next section, but we focus on the large gap case in this section.

Given this approximation that $\der{C_0}{s} = 0$, we assume that it is a good approximation that $C_0(s) = 1$ for the entire evolution.  Essentially we are saying that the majority of the amplitude remains in the ground state with minimal changes to its value.  With this assumption, Eq.~\ref{eq:C1eq1} becomes
\begin{equation}
      i\der{C_1}{s} - i \frac{\gamma}{\Delta} = \tau\Delta C_1.
\end{equation}
Notice that $C_1$ is kept in this equation despite being disregarded in the $C_0$ equation.  It is kept both because we are now looking at the change in $C_1$ itself and because this otherwise small term is multiplied by $\tau$ which is taken to be large in the near-adiabatic limit.  This differential equation has an integral solution when $C_1(s=0) = 0$
\begin{equation}
      C_1(s, \tau) = \int_0^s dx \frac{ e^{-i \tau  \int_{x}^s dz\, \Delta(z) }}{\Delta(x)/\gamma(x)}.
\end{equation}

What we are actually going to care about is the final amplitude after the total evolution time $\tau$, so the quantity we work with is mainly
\begin{equation}
      \label{eq:integral}
      C_1(1, \tau) = \int_0^1 ds \frac{ e^{-i \tau  \int_{s}^1 dz\, \Delta(z) }}{\Delta(s)/\gamma(s)}
\end{equation}

Our goal is to approximate this integral in the limit of large $\tau$ so that we can find the probability amplitude for leaving the ground state and entering one of these excited states.

Throughout this approximation, we need to assume that $\Delta(s)$ does not become exceedingly small.  We can rewrite our integral as
\begin{equation}
      \label{eq:integral2}
      C_1(1, \tau) = \int_0^1 ds \frac{\der{}{s}e^{-i \tau  \int_{s}^1 dz\, \Delta(z) }}{i\tau \Delta(s)^2/\gamma(s)}
\end{equation}
Integration by parts yields
\begin{align}
      \label{eq:integral3}
      C_1(1, \tau) &= \left[\frac{e^{-i\tau\int_s^1 dz\, \Delta(z)}}{i\tau\Delta(s)^2/\gamma(s)}\right]_{s=0}^1\\
      \nonumber&-\int_0^1 ds \,e^{-i \tau  \int_{s}^1 dz\, \Delta(z) }\der{}{s}\frac{1}{i\tau \Delta(s)^2/\gamma(s)}.
\end{align}
By the properties of oscillatory integrals, the last integral here is $\mathcal{O}(\tau^{-2})$, so we are left with
\begin{equation}
      \label{eq:integral4}
      C_1(1, \tau) = \frac{1}{i\tau\Delta(1)^2/\gamma(1)}-\frac{e^{-i\tau\int_0^1 ds\, \Delta(s)}}{i\tau\Delta(0)^2/\gamma(0)}+\mathcal{O}(\tau^{-2}).
\end{equation}
For convenience, we will define $\rho(s) = \gamma(s)/\Delta(s)^2$, and this value $\rho(s)$ is related to the naive adiabatic condition that $\tau\gg \int_0^1ds\,|\rho(s)|$.  Therefore, this $\rho(s)$ can be thought of as the gauge by which we can determine whether we are in the adiabatic limit.

Then the probability of transitioning into one of the $m$-fold degenerate first excited states is given by
\begin{align}
      \label{eq:prob_fail_large_gap}
      \frac{P(\tau)}{m} &= \frac{\rho(1)^2+\rho(0)^2}{\tau^2}\\
                        &-  \frac{2\rho(0)\rho(1)}{\tau^2}\cos(\omega\tau) + \mathcal{O}(\tau^{-3}),\nonumber
\end{align}
where
\begin{equation}
      \label{eq:large_omega}
      \omega \equiv \int_0^1 ds\, \Delta(s).
\end{equation}

Therefore, the final probability of failure (and success) are oscillating functions with a frequency dependent on the integral of the spectral gap.  This result is a previously known result \cite{Lidar,Grandi, Wiebe}.  Additionally, Wiebe and Babcock \cite{Wiebe} have proposed using this oscillating behavior to enhance quantum adiabatic computing.

Notice that the rule of thumb for the adiabatic condition states that the evolution time, $\tau$, needs to grow with $\rho(s)$, and indeed our formula captures this since the probability of failure depends on $\frac{\rho(s)}{\tau}$.  The new and interesting behavior here is not the overall $\tau$ dependence but the oscillating dependence.

\subsection{Numerical Confirmation}
\label{ssec:large_num}

\begin{figure}
      \includegraphics[width=0.48\textwidth]{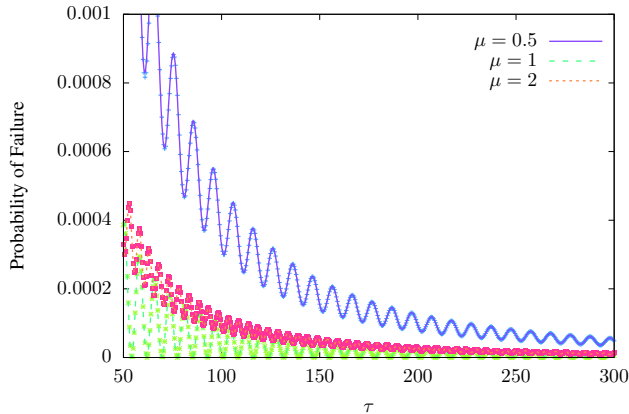}
      \caption{            
                 The probability of transition to the first excited state versus the evolution time $\tau$ for no barrier.  The solid lines represent the theoretical predictions coming from Eq.~\ref{eq:prob_fail_large_gap}, and the circles represent data obtained from direct integration of the Schr\"odinger equation.  Data is shown for various $\mu$ and $n=1$; though this problem has decoupled qubits, so this is representative of arbitrary $n$.  Notice especially the oscillatory behavior that depends on the integral of the spectral gap over the entire evolution.
      }
      \label{fig:no_bump}
\end{figure}

\begin{figure}
      \includegraphics[width=0.48\textwidth]{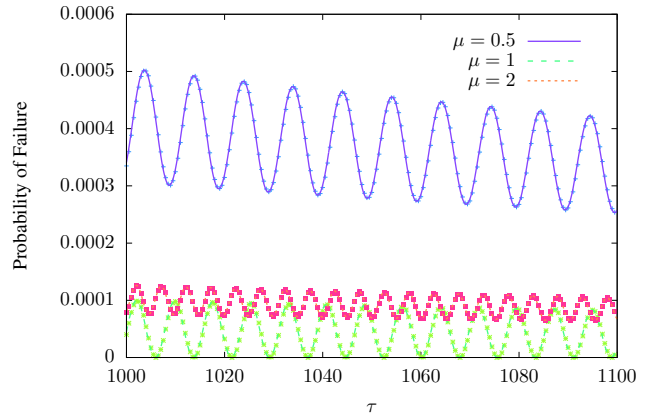}
      \caption{            
                 The probability of transition to the first excited state versus the evolution time $\tau$ for a barrier that grows with $\alpha=\beta=1/10$.  The solid lines represent the theoretical predictions coming from Eq.~\ref{eq:prob_fail_large_gap}, and the circles represent data obtained from direct integration of the Schr\"odinger equation.  Data is shown for various $\mu$ and $n=100$.  Even in this case with more approximations than in Fig.~\ref{fig:no_bump}, the oscillatory analytic prediction is still quite accurate.
      }
      \label{fig:large_gap_bump}
\end{figure}

To numerically test the predictions of the previous subsection, we return to the quantum computational problem of Hamming weight barrier tunneling introduced in subsection \ref{ssec:QAO}.

One important and instructive way to get a large gap out of the barrier tunneling problem is to set the barrier to zero, $b(|z|)=0$.  This decouples all the qubits from each other, making the problem effectively a two-level system, independent of the number of qubits.  As well, the spectral gap has a simple closed form expression for this case
\begin{equation}
      \label{eq:no_barrier_gap}
      \Delta_{NB}(s) = \sqrt{1-2s+(1+\mu)s^2}
\end{equation}
This gap is also useful even in the cases with a barrier because it can well approximate the gap far away from where the tunneling event occurs.  Also note that in the no barrier case $\gamma(s) = \frac{\mu}{2\Delta(s)}$.  In later analyses of cases with barriers, we use this expression for $\gamma(s)$ again when evaluating $\gamma(0)$ and $\gamma(1)$ which are far from the tunneling event.

When there is no barrier, we can analytically integrate Eq.~\ref{eq:no_barrier_gap} and obtain a form for the probability of failure, Eq.~\ref{eq:prob_fail_large_gap}.  In Fig.~\ref{fig:no_bump}, we compare this analytic expression for the near-adiabatic probability of transitioning to an excited state with the exact result obtained by numerical integration of the Schr\"odinger equation.  The analytic approximations predicts the actual data extremely well.  It should be noted that without a barrier, the qubits are decoupled, so this is a two-level system, meaning many of our approximations are exact.

In Fig.~\ref{fig:large_gap_bump}, we look at a case, where the barrier is present and the qubits are not decoupled.  In this case, we take a barrier with scaling exponents $\alpha=\beta=1/10$.  Based on previous work \cite{Reichardt}, this barrier should be easy to tunnel through, with only a constant gap as $n$ increases.  In this instance, we have taken $n=100$ qubits and see that good agreement between the direct data and the analytic approximation from Eq.~\ref{eq:prob_fail_large_gap} that was calculated using numerical integration of the spectral gap and approximating $\rho(1)$ and $\rho(0)$ by the unperturbed value since they are evaluated far from where the barrier is relevant.

\section{Small Gap}
\label{sec:small_gap}

For the small gaps, we examine the case where the gap remains large everywhere except in a region right around a critical $s^*$.  At this critical $s^*$, the system has an avoided level crossing, which traditionally is handled by a formalism such as the Landau-Zener problem.  In this section, we focus on a system with a single avoided level-crossing, but our methods can easily be generalized to systems with multiple level crossings.

Note that this section discusses the near-adiabatic limit, so big-O notation is not appropriate here.  In the $\tau\to\infty$ limit, the results of the large gap section are accurate.  This section focuses on the behavior of the success probability in regions where the inverse gap is small relative to the evolution-time, $\tau$.  Thus, we examine an intermediate region, and all of our results for modifications to $C_1(1,\tau)$ tend to zero faster than Eq.~\ref{eq:integral4} in the asymptotic limit of $\tau$.

\subsection{Frequency Splitting}
\label{ssec:freq_splitting}

An integral of the form Eq.~\ref{eq:integral} can often be treated with the stationary phase approximation.  If the phase function, $\int_s^1 dz\,\Delta(z)$, is ever stationary as would occur when $\Delta(s)=0$, then the stationary phase approximation says that asymptotically, the integral is dominated by the value close to that stationary point.

Unfortunately, since the gap never goes to zero, $\Delta(s)\neq 0$, we never have a point of true stationary phase.  However, we have an avoided level crossing where the gap becomes very small in the vicinity of $s^*$.  Additionally, near $s^*$, the denominator of the integral is also small since it is proportional to $\Delta(s)$ ($\gamma(s)$ also often depends inversely on the gap).
Therefore, the region around $s^*$ should still contribute more to the integral than other regions, but since this is not a true stationary phase point, the contribution near $s^*$ is drowned out in the asymptotic limit of $\tau$.

Therefore, we will make an ansatz that the region around $s^*$ also contributes significantly to the final probability amplitude in the near-adiabatic limit.  The contribution to the probability amplitude in the vicinity of $s^*$ is roughly of the form
\begin{equation}
      \label{eq:small_con}
      \lambda \equiv e^{-i \tau  \int_{s^*}^1 dz\, \Delta(z) }\int_{s^*-\epsilon}^{s^*+\epsilon} ds \frac{ e^{-i \tau  \int_{s}^{s^*} dz\, \Delta(z) }}{\Delta(s)/\gamma(s)},
\end{equation}
where we have pulled out the contribution to the phase due to getting to the critical point.  We can generalize this further (potentially allowing us to relax some of the assumptions that led to Eq.~\ref{eq:integral}) to
\begin{equation}
      \label{eq:small_con2}
      \lambda = e^{-i \tau  \int_{s^*}^1 dz\, \Delta(z)}\Lambda(\tau).
\end{equation}
Later in this section we present numeric evidence supporting this ansatz.  Furthermore, our numerics indicate that $\Lambda(\tau)$ is real, allowing us to ignore any potential extra phases.

Then, our conjectured probability amplitude in the near-adiabatic limit is
\begin{equation}
      \label{eq:small_con3}
      C_1(1, \tau) \approx \frac{\rho(1)}{i\tau}-\frac{\rho(0)e^{-i\tau(\omega_++\omega_-)}}{i\tau}+\Lambda(\tau)e^{-i \tau  \omega_+},
\end{equation}
\begin{equation*}
      \omega_+ = \int_{s^*}^1 dz\, \Delta(z),~~~~~\omega_- = \int_0^{s^*} dz\, \Delta(z).
\end{equation*}

This probability amplitude leads to a probability of transition of

\begin{align}
      \label{eq:freq_split_prob}
      \frac{P(\tau)}{m} &\approx \Lambda(\tau)^2+\frac{\rho(0)^2+\rho(1)^2}{\tau^2}\\
              &  +  \frac{2\Lambda(\tau)}{\tau} \left(\rho(0)\sin(\omega_- \tau)+\rho(1)\sin(\omega_+\tau)\right)\nonumber\\
              &  -  \frac{2\rho(0)\rho(1)}{\tau^2}\cos((\omega_++\omega_-)\tau)\nonumber
\end{align}
where $m$ is the degeneracy of the first excited state.

Most importantly the final probability amplitude now has sinusoidal motion dependent on two frequencies, $\omega_\pm$.  Thus, the final probability no longer has a simple sinusoidal behavior but depends on the superposition of multiple sinusoids.

This splitting of the frequency, creating a superposition of sinusoids when the gap is small, is a well realized feature in actual problems, as seen numerically in Section~\ref{ssec:small_num}.  In fact, this frequency splitting seems to persist even when most of the simplifying assumptions that went into Eqs.~\ref{eq:prob_fail_large_gap} \& \ref{eq:freq_split_prob} fail.  In the next section, we take a Landau-Zener approach to the small gap, but in other small gap models we examined, this frequency splitting persisted.

\subsection{Analytic Approximation}
\label{ssec:small_anal}

In this section, we approximate our avoided level crossing as a Landau-Zener transition.  In the language of the previous subsection, we use the ansatz that $\Lambda(\tau)$ is related to the Landau-Zener transition probability.  The Landau-Zener problem works with Hamiltonians of the form
\begin{equation}
      \label{eq:ham_LZ}
      \hat{H}_{LZ}(s) = \frac{v}{2}(s-s^*) \hat{\sigma}_z+\frac{g}{2}\hat{\sigma}_x,
\end{equation}
where $g \equiv \Delta(s^*)$ is the minimum spectral gap, and $v$ is slope of the spectral gap far from $s^*$.  The Landau-Zener formula says that the probability of transitioning from the ground state to the excited state going from $t=-\infty$ to $t=\infty$ is
\begin{equation}
      P_{LZ} = e^{-2\pi \frac{g^2}{4v}\tau},
\end{equation}
where the $\tau$ is coming in because the Landau-Zener transition is formulated in actual time, $t$, whereas Eq.~\ref{eq:ham_LZ} is formulated in $s$.

We take the probability amplitude of transition through our avoided level crossing to be proportional to $\sqrt{P_{LZ}}$.  We also include a real parameter, $A$, to account for non-idealnesses in the Landau-Zener transition such as the finite nature of our transition and the fact that we do not start the avoided level-crossing in exactly the ground state.  Therefore, we take
\begin{equation}
      \Lambda(\tau) = A e^{-\pi \frac{g^2}{4v}\tau}.
\end{equation}

Then, our ansatz for the final failure probability when there is an avoided-level crossing is
\begin{align}
      \label{eq:freq_split_prob_LZ}
      \frac{P(\tau)}{m} &\approx A^2 e^{-2\pi \frac{g^2}{4v}\tau}+\frac{\rho(0)^2+\rho(1)^2}{\tau^2}\\
              &  +  \frac{2A e^{-\pi \frac{g^2}{4v}\tau}}{\tau} \left(\rho(0)\sin(\omega_- \tau)+\rho(1)\sin(\omega_+\tau)\right)\nonumber\\
              &  -  \frac{2\rho(0)\rho(1)}{\tau^2}\cos((\omega_++\omega_-)\tau).\nonumber
\end{align}

In an actual setting, $g$, $v$, and $\omega_\pm$ can be determined from the shape of the spectral gap of the problem in question.  We leave $A$ as a fitted parameter that accounts for non-idealness in our system.  In the next section, we determine all these parameters in specific computational settings.

\subsection{Numerical Confirmation}
\label{ssec:small_num}

\begin{figure}
      \includegraphics[width=0.48\textwidth]{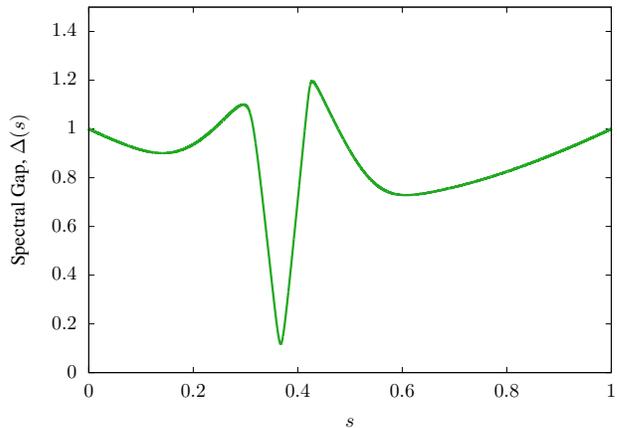}
      \caption{            
                 The spectral gap, $\Delta(s)$, versus $s$, for a binomial barrier with $\alpha=0.3$, $\beta=0.5$, $\mu=1$, and $n=84$.  The minimum spectral gap, $g$ is obtained from the minimum here, and the Landau-Zener slope, $v$, is approximated by the almost linear sections near the minimum gap.  As well the frequencies, $\omega_\pm$, come from numerical integration of this curve.
      }
      \label{fig:ab35mu1gap}
\end{figure}

\begin{figure}
      \includegraphics[width=0.48\textwidth]{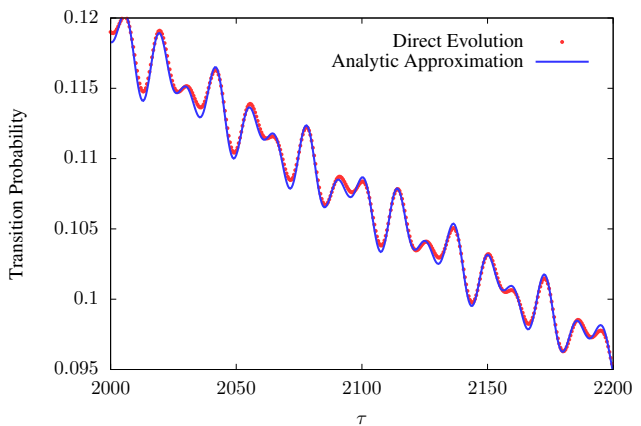}
      \caption{            
                 The probability of transitioning out of the ground state as a function of the total evolution or run time, $\tau$, for a binomial barrier with $\alpha=0.3$, $\beta=0.5$, $\mu=1$, and $n=84$.  The red dots represent data obtained through direct evolution of the Schr\"odinger equation, and the blue curve is the result of applying Eq.~\ref{eq:freq_split_prob_LZ} to the problem.  To obtain the blue curve, the parameter $A$ was fitted to $A=0.107$, but all other parameters were calculated from the spectral gap directly.  Notice that the analytic expression matches up quite well with the data, mimicing the frequency behavior.  Also notice that this analytic expression seems to hold even when the probability of failure is relatively high, around $10\%$.
      }
      \label{fig:ab35mu1}
\end{figure}

%\begin{figure}
%      \includegraphics[width=0.48\textwidth]{a=0-3_b=0-5_mu=1_n=84_close.eps}
%      \caption{            
%                 This plot is just a zoomed in version of Fig.~\ref{fig:ab35mu1}.
%      }
%      \label{fig:ab35mu1_close}
%\end{figure}

\begin{figure}
      \includegraphics[width=0.48\textwidth]{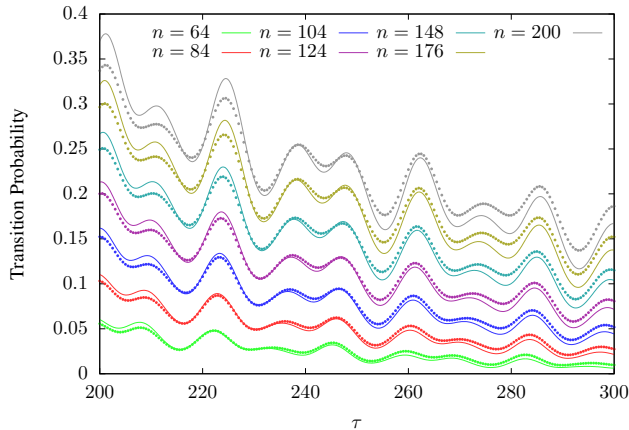}
      \caption{            
                 The probability of transitioning out of the ground state versus the evolution time, $\tau$, compared to the theoretical predictions from Eq.~\ref{eq:freq_split_prob_LZ}.  The dots are the direct Schr\"odinger evolution data, and the lines are the thoeretical predictions.  This data all comes from the Hamming weight barrier problem with a binomial barrier with ehight and width both growing with $n^{0.3}$ and $\mu=1$.  The frequency behavior seems to be captured by the theoretic predictions well, but the overall amplitude is less well-predicted especially for higher $n$ where the probability of trnasitioning is higher.
      }
      \label{fig:0-3_many_n}
\end{figure}

In this section, we present numeric data confirming the usefulness of the analytics in the rest of the section.  Our approximations rely on two key assumptions, namely that $|C_0(s)|$ is close to one, meaning we primarily stay in the grounds state and that $\tau$ is large.  We will test these approximations as well as the Landau-Zener anstaz modification by comparing Eq.~\ref{eq:freq_split_prob_LZ} to direct data.

Each numeric simulation is based on direct Schr\"odinger evolution of the wavefunction from $s=0$ to $s=1$.  This evolution then gives us data of $P(\tau)$ (actually we calculate $1-P(\tau)$, the probability of staying in the ground state) versus $\tau$.  This data is then fit using a function of the form of Eq.~\ref{eq:freq_split_prob_LZ} to obtain a fitted value of $A$.  Across numerous trials with different barrier shapes and sizes, we determine that $A$ can be taken as real.

For each of our simulations, we numerically calculate the gap as a function of $s$ and use it to extract $g$, $v$, and $\omega_\pm$.  For $v$, we base its value on the slope of $\Delta(s)$ close to the avoided level crossing where the gap is increasing or decreasing effectively linearly.  The frequencies $\omega_\pm$ are obtained through numerical integration of the spectral gap before and after the critical $s^*$.

We have done several simulations for the Hamming weight barrier problem with a binomial barrier.  For a representative plot, see Fig.~\ref{fig:ab35mu1} which shows the close correspondence between the analytic expression in Eq.~\ref{eq:freq_split_prob_LZ} and the direct data, for a binomial barrier with height and width scaling like $n^{0.5}$ and $n^{0.3}$ respectively and with $\mu=1$ and $n=84$.  Notably, the frequency splitting behavior is evident here in the superposition of two sinusoids, and the overall scaling matches well with the Landau-Zener probability of transition exponential.

The fitted parameter $A$ in Fig.~\ref{fig:ab35mu1} is $A=0.107$.  In a pure Landau-Zener transition, the value of $A$ is $1$; the fact that our value is less than $1$, is likely an indicator of the finite scale of our Landau-Zener region.  A true Landau-Zener transition occurs from $t=-\infty$ to $t=\infty$, so we have a finite range, which might influence the value of $A$.  Also the $A$ parameter is probably absorbing discrepancies caused by our other assumptions in deriving Eq.~\ref{eq:freq_split_prob_LZ}, including those independent of the Landau-Zener-like transition.

We performed similar trials for other barrier sizes and values of $n$ and $\mu$, and for each trial, we found a very good correspondence between our anstaz and the direct Schr\"odinger data.  One of the largest assumptions we use in deriving Eq.~\ref{eq:freq_split_prob_LZ} is that the majority of the probability remains in the ground state so that we can approximate $C_0(s)\approx 1$ in our differential equations.  Our numerics show that this is somewhat of a loose condition with, for instance, Fig.~\ref{fig:ab35mu1} showing good correspondance even when $|C_0(s)|^2\approx 0.9$.

Additionally, our approximations also rely on the fact that $\tau$ is large.  In Fig.~\ref{fig:0-3_many_n} we display data compared to predictions for a variety of $n$ values at much lower $\tau$.  The agreement is not as clear as with larger $\tau$, and especially at larger $n$ (thus larger transition probability), the agreement is noticeably degraded.  However, there is still correspondence, and especially the frequencies, if not the amplitudes line up well.

In Fig.~\ref{fig:0-3_many_n}, all this data is taken for a barrier with height and width scaling with $n^{0.3}$.  The spectral gap $\Delta(s)$ is very similar for all these $n$ values.  As $n$ increases, the gap around the avoided level-crossing changes, lowering $g$ and raising $v$, but the majority of the spectral gap remains the same, meaning that $\omega_\pm$ are virtually the same for different $n$.  The persistence of the same $\omega_\pm$ leads to similar frequency behavior across $n$, even though the enveloping probability scaling with $\tau$ changes with $g$, in accordance with the standard adiabatic theorem.

\begin{figure}
      \includegraphics[width=0.48\textwidth]{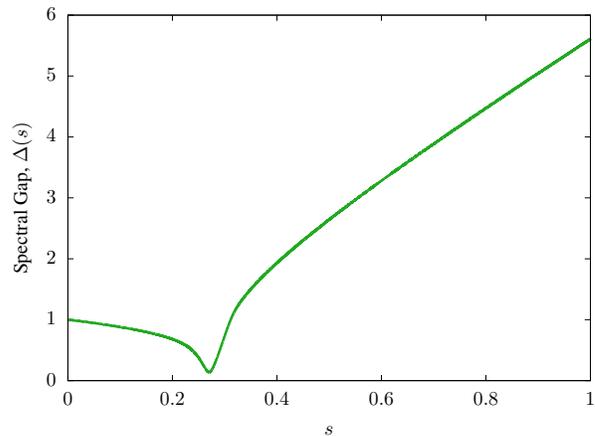}
      \caption{            
                 The spectral gap, $\Delta(s)$, versus $s$, for the cubic potential and $n=30$.  The frequencies, $\omega_\pm$, come from numerical integration of this curve.  Notice that the avoided level crossing is assymmetric for $n=30$, leaving us unable to properly use the Landau-Zener transition probability to approximate this crossing.
      }
      \label{fig:cubic_gap}
\end{figure}

\begin{figure}
      \includegraphics[width=0.48\textwidth]{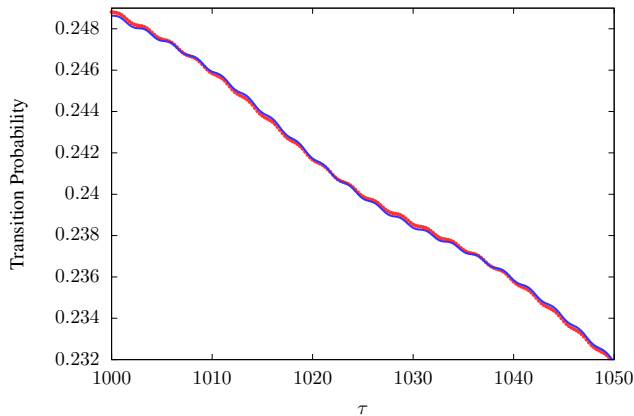}
      \caption{            
                 The probability of transitioning out of the ground state as a function of the total evolution or run time, $\tau$, for the cubic potential and $n=30$.  The red dots represent data obtained through direct evolution of the Schr\"odinger equation, and the blue curve is the result of applying Eq.~\ref{eq:freq_split_prob_LZ} to the problem.  Due to the assymetric gap during the avoided level crossing as seen in Fig.~\ref{fig:cubic_gap}, the fit needed more than just $A$ as a fitted parameter.  We also included $v$ as a fitted parameter to obtain this agreement.  Notice that the oscillatory pattern is still correct, confirming the robustness of the frequency splitting behavior from section \ref{ssec:freq_splitting}.
      }
      \label{fig:cubic_evolution}
\end{figure}

For the cubic potential, the spectral gap still goes through an avoided level crossing as shown in Fig.~\ref{fig:cubic_gap}, but in this case, the avoided level crossing is much more assymetric for $n=30$.  The slopes on either side of the avoided level crossing are different, making it difficult to determine $v$.  Therefore, in this case, we leave $v$ as a fitted parameter.  This essentially eliminates the benefit of our Landau-Zener ansatz, but the frequency splitting of Eq.~\ref{eq:freq_split_prob} is still valid.  We also calculate $\rho(0)$ and $\rho(1)$ numerically using equivalent diagonalization methods to how we calculate the spectral gap itself.

Using Eq.~\ref{eq:freq_split_prob_LZ} with fitted $A$ and $v$, we predict how the transition probability changes as a function of $\tau$ in Fig.~\ref{fig:cubic_evolution}.  The correspondance between our prediction and the direct evolution data is quite good, but much of this is due to two fitted parameters.  The important part of this figure is that the oscillatory behavior matches quite well between the predicition and direct data.  

Also, Fig.~\ref{fig:cubic_evolution} once again matches well even at the relatively high transition probability of around $0.24$, indicating that the frequency splitting behavior is fairly robust to our exact approximations.  The correspondance in this cubic potential problem as well as the barrier problem indicate the correctness of our frequency splitting ansatz.

% Cite "Tight Bounds on Quantum Searching" for the digital Grover periodic behavior

\section{Adiabatic Grover Search}
\label{sec:grover}

The Grover search algorithm \cite{Grover,Boyer} is a digital quantum algorithm for searching an unstructured set of $N$ elements for one of $M$ target elements, where $M\ll N$.  Classically, $\mathcal{O}(N/M)$ queries must be made to find one of the target states, but Grover's algorithm requires only $\mathcal{O}(\sqrt{N/M})$ oracular calls to find a target with high probability \cite{Boyer}.  Notably, the likelihood of success of this algorithm is periodic in the number of queries with period $\propto\sqrt{N/M}$.  Therefore, timing needs to be exact to achieve success; though, there are methods of alleviating this periodicity in favor of a larger constant scaling factor.

An adiabatic version of Grover's search algorithm has been developed \cite{Roland,vanDam02} that mimics the $\mathcal{O}(\sqrt{N/M})$ scaling.  However, previous studies of this algorithm have relied on studying the adiabatic theorem's asymptotic scaling, and to our knowledge, no groups have looked into whether the periodic nature of digital quantum search carries over to adiabatic Grover in some way.  In this section, we demonstrate that the Grover oscillations exist in adiabatic search and are a direct result of the large gap oscillations described in section \ref{sec:large_gap}.

\subsection{Adiabatic Grover Background}

The adiabatic grover algorithm is setup on a Hilbert space with dimension $N$ (not to be confused with $n$ qubits discussed in previous sections).  The initial Hamiltonian is just a simple connection Hamiltonian between all basis states:
\begin{equation}
      \hat{H}_0 = -\frac{1}{N} \sum_{i,j=1}^N \ket{i}\bra{j}.
\end{equation}
The ground state of this Hamiltonian is just the uniform superposition of all states.  For ease of analysis, typically the first $M$ basis states are chosen to be the target states, and the final Hamiltonian gives them a preferential energy term such that
\begin{equation}
      \hat{H}_1 = \hat{\mathbb{I}}-\sum_{m=1}^M \ket{m}\bra{m}.
\end{equation}

Note that the ground state of the final Hamiltonian is $M$-fold degenerate; whereas, the ground state of the initial Hamiltonian is non-degenerate.  Seemingly, this means that the spectral gap would go to zero at some point during the evolution; however, this is not an issue due to the nature of the degeneracy.  The true ground state throughout the evolution is symmetric between target states so that in the end it is a uniform superposition of all targets.  The degeneracy in the final ground state is due to states that are non-symmetric in target states transitioning between the first excited state and the ground state.  Since the Grover Hamiltonian and initial ground state are symmetric between target states, and since we are concerned with coherent evolution, these non-symmetric states are inaccessible, meaning that we can largely ignore them.

A linear interpolation between these two Hamiltonians does not lead to the square root speedup of Grover's algorithm \cite{Roland}.  Instead, a more generalized interpolation needs to be considered
\begin{equation}
      \hat{H}(s) = (1-g(s))\hat{H}_0 + g(s) \hat{H}_1.
\end{equation}
The spectral gap for this Hamiltonian is given by
\begin{equation}
      \label{eq:grover_gap}
      \Delta(s) = \sqrt{1-4 (1-g(s)) g(s) \frac{(N-M)}{N}}.
\end{equation}

The optimal annealing schedule \cite{Roland}, utilizes the adiabatic condition, Eq.~\ref{eq:adi_condition}, slowing down when the spectral gap is small and speeding up when it is large.  By optimizing for the amount of time spent in the small region, the ideal annealing schedule uses
\begin{equation}
      \label{eq:grover_schedule}
      g(s) = 
      \frac{1}{2} \left(1-
      \frac{\tan \left((1-2 s) \tan ^{-1}\left(\sqrt{\frac{N-M}{M}}\right)\right)}
      {\sqrt{\frac{N-M}{M}}}\right)
\end{equation}

This optimal annealing schedule leads to an adiabatic runtime that is $\mathcal{O}\left(\sqrt{N/M}\right)$.

\subsection{Large Gap Oscillations}

\begin{figure}
      \includegraphics[width=0.48\textwidth]{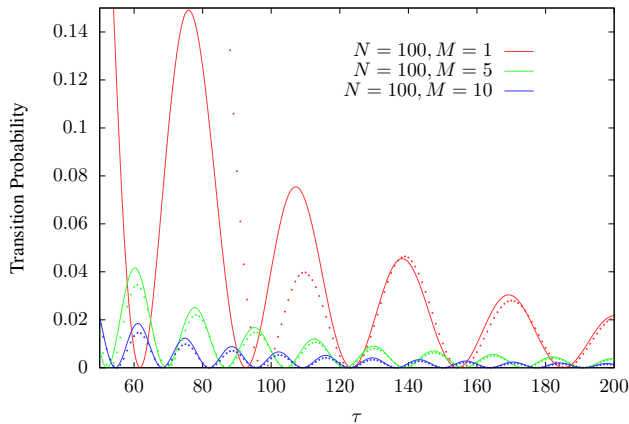}
      \caption{            
                 The probability of transitioning out of the ground state as a function of the total evolution or run time, $\tau$, for the adiabatic Grover search problem.  Notice that there is good agreement between the analytics and the data when the probability of transitioning is low but that the agreement is worse for shorter evolution times.  The analytics here are exact and included no fitting parameters.
      }
      \label{fig:grover}
\end{figure}

Our contribution is to show that the large gap oscillations described in section \ref{sec:large_gap} lead to the same periodic behavior in the adiabatic algorithm as in the standard digital Grover algorithm.  Note that the gap in the Grover problem does become small at $s=1/2$, where $\Delta(1/2) = \sqrt{M/N}$, but in this section we consider only the large gap oscillations.  Also, Wiebe and Babcock \cite{Wiebe}, have examined the large gap oscillations of adiabatic Grover in the case of $M=1$.  There, they show that timing the adiabatic algorithm based on the large gap oscillations improves the performance of the algorithm; however, they never explicitly state the oscillation frequency or connect it back to digital Grover.  Furthermore, our results make the simple extension to $M>1$.

Looking at Eq.~\ref{eq:freq_split_prob_LZ}, it is easy to see that the large gap oscillations will dominate if the standard adiabatic condition, Eq.~\ref{eq:adi_condition}, is met.  Therefore, in this section the system is evolving adiabatically so that $\tau\in\mathcal{O}\left(\sqrt{N/M}\right)$, and our main goal is to determine what the fine structure of the oscillations in this limit are.

Using the spectral gap and the annealing schedule, Eqs.~\ref{eq:grover_gap} \& \ref{eq:grover_schedule}, we can calculate the frequency of the large gap oscillations as described by Eq.~\ref{eq:large_omega}:
\begin{equation}
      \omega = \sqrt{\frac{M}{N}} \frac{\tanh^{-1}\sqrt{\frac{N-M}{N}}}{\tan^{-1}\sqrt{\frac{N-M}{M}}}
\end{equation}
The period of oscillations $T\equiv 1/\omega$ in the limit of large $N/M$ is given by
\begin{equation}
      T \to \frac{\pi}{2\ln 2+\ln \frac{N}{M}}\sqrt{\frac{N}{M}}.
\end{equation}
Therefore, up to logarithmic factors, the period of oscillations for adiabatic Grover's search is $\widetilde{\mathcal{O}}(\sqrt{\frac{N}{M}})$ which matches the digital equivalent.

Additionally, we can look at the amplitudes of the oscillations in the probability of transitioning, Eq.~\ref{eq:prob_fail_large_gap}.  In order to get $\rho(s)$, we need $\Delta(s)$ which was listed in the previous subsection and $\gamma(s)$.  Assuming the annealing schedule in Eq.~\ref{eq:grover_gap}
\begin{align}
      \gamma(s) &= 4 M \tan ^{-1}\left(\sqrt{\frac{N-M}{M}}\right)\\\nonumber
                &\times\frac{ \left(\sqrt{\frac{N-M}{M}}-\tan \left((1-2 s) \tan ^{-1}\left(\sqrt{\frac{N-M}{M}}\right)\right)\right)}
      {N-2M+N \cos \left(2 (1-2 s) \tan ^{-1}\left(\sqrt{\frac{N-M}{M}}\right)\right)}.
\end{align}

The truly important part of this equation is that this $\gamma(s)$ is symmetric about $s=1/2$ so that notably $\gamma(0) = \gamma(1)$.  Since the spectral gap is $\Delta(0) = \Delta(1) = 1$ at the end points as well, this leaves us with $\rho \equiv \rho(0) = \rho(1)$.  Therefore, the probability of transitioning out of the ground state reduces to
\begin{equation}
      \label{eq:grover_pred}
      P(\tau) = \frac{4\rho^2}{\tau^2}\sin^2\left(\frac{\omega\tau}{2}\right) + \mathcal{O}(\tau^{-3}).
\end{equation}

Notably, this probability goes to $\mathcal{O}(\tau^{-3})$ periodically according to $\omega$.  Therefore, adiabatic Grover's search can be timed to get perfect success probability in the adiabatic limit.  In Fig.~\ref{fig:grover}, we show the agreement between the analytic predictions of Eq.~\ref{eq:grover_pred} and direct evolution data.  There is good agreement between the analytics and the data, especially for lower transition probabilities.

\section{Conclusion}
\label{sec:conc}

We have worked in the near-adiabatic limit, expanding upon the evolution time dependence.  Specifically we have explored how the probability of transitioning out of the ground state depends on $\tau$, the total evolution time.

In the adiabatic limit, the probability of transition decreases with $\tau^{-2}$, but we explored the structure on top of this basic decay, looking at the sinusoidal behavior superposed on top.  In the absence of a small gap, the sinusoidal behavior has a frequency that depends only the integral of the spectral gap along the evolution.  When an avoided level crossings occur, the sinusoidal behavior becomes the superposition of frequencies that depend on the integrals of the spectral gap, broken up at the avoided crossings.

We back up our analytics with numerics that very closely match up with our predictions.  Specifically our numerics are in the context of quantum computing and annealing, where this work can predict how long to run the algorithm to lead to oscillatory enhancements to the success probability.

Our work on frequency splitting in Sec.~\ref{ssec:freq_splitting} relies largely on an ansatz inspired by the stationary phase approximation and the Landau-Zener transition.  Further work could be done to remove the need for an ansatz here, working more from first principles.  The fact that the numerics match the analytics even when the approximations are less well-founded, indicates that stronger analytic work could be possible.

\subsubsection*{Acknowledgements}
This material is based upon work supported by the National Science Foundation under Grant No. 1620843.

%%%%%%%%%%%%%%%%%%%%%%%%%%%%%%%%%%%%%%%%%%%%%%%%%%%%%%%%%%%%%%%%%%%%%%%%%


\begin{thebibliography}{99}

\bibitem{BornFock} M. Born, V. Fock, Zeitschrift f\"ur Physik {\bf 51} (3-4) 165–180 (1928).

\bibitem{Farhi2000} E. Farhi, J. Goldstone, S. Gutmann, M. Sipser, quant-ph/0001106 (2000).

\bibitem{Aharonov} D. Aharonov, W. van Dam, J. Kempe, Z. Landau, S. Llyod, O. Regev, SIAM J. Comp. {\bf 37}, 166 (2007).

\bibitem{Farhi2002} E. Farhi, J. Goldstone, S. Gutmann, quant-ph/0201031 (2002).

\bibitem{Farhi2008} E. Farhi, J. Goldstone, S. Gutmann, D. Nagaj. Int. J. Quantum Inf. {\bf 6}, 3 (2008).

\bibitem{Martonak} R. Marto\u{n}\'{a}k, G. E. Santoro, E. Tosatti. Phys. Rev. B {\bf66}, 094203 (2002).

\bibitem{Hastings} M. B. Hastings, M. H. Freedman, Quant. Inf. \& Comp. {\bf 13}, 11-12 (2013).

\bibitem{Boxio2} S. Boixo, T. F. R{\o}nnow, S. V. Isakov, Z. Wang, D. Wecker, D. A. Lidar, J. M. Martinis, M. Troyer, Nature Phys. {\bf10}, 218 (2014).

\bibitem{Battaglia} D. Battaglia, G. Santoro, E. Tosatti, Phys. Rev. E {\bf71}, 066707 (2005).

\bibitem{Heim} B. Heim, T. F. R{\o}nnow, S. V. Isakov, M. Troyer, Science {\bf 348}, 6231 (2015).

\bibitem{Crosson} E. Crosson, M. Deng, quant-ph/1410.8484 (2014).

\bibitem{Harrow} E. Crosson, A. Harrow, Proc. of FOCS 2016, pp. 714--723 (2016).

\bibitem{Muthukrishnan} S. Muthukrishnan, T. Albash, D. A. Lidar, Phys. Rev. X {\bf 6}, 031010 (2016).

\bibitem{Brady} L. Brady, W. van Dam, Phys. Rev. A {\bf 93}, 032304 (2016).

\bibitem{Jiang}  Z. Jiang, V. N. Smelyanskiy, S. V. Isakov, S. Boixo, G. Mazzola, M. Troyer, and H. Neven, quant-ph/1603.01293 (2016).
(2016).

\bibitem{Kong} L. Kong, E. Crosson, quant-ph/1511.06991 (2015).

\bibitem{DWave}  M. W. Johnson, M. H. S. Amin, S. Gildert, T. Lanting, F. Hamze, N. Dickson, R. Harris, A. J. Berkley, J. Johansson, P. Bunyk, E. M. Chapple, C. Enderud, J. P. Hilton, K. Karimi, E. Ladizinsky, N. Ladizinsky, T. Oh, I. Perminov, C. Rich, M. C. Thom, E. Tolkacheva, C. J. S. Truncik, S. Uchaikin, J. Wang, B. Wilson, G. Rose. Nature {\bf 473}, 7346 (2011).

\bibitem{Brady3} L. Brady, W. van Dam, Phys. Rev. A {\bf 95}, 032335 (2017).

\bibitem{Lidar} D. A. Lidar, A. T. Rezakhani, A. Hamma, J. Math. Phys. {\bf 50}, 102106 (2009).

\bibitem{Grandi} C. De Grandi, A. Polkovnikov, in \emph{Quantum Quenching, Annealing and Computation}, edited by A. Das, A. Chandra and B. K. Chakrabarti (Springer, Heidelberg 2010), Vol. 802 pp. 75--114. 

\bibitem{Wiebe} N. Wiebe, N. S. Babcock, New J. Phys. {\bf 14}, 013024 (2012).

\bibitem{Landau} L. Landau, Physikalische Zeitschrift der Sowjetunion {\bf 2}, pp. 46--51 (1932).

\bibitem{Zener} C. Zener, Proc. of the Royal Society of London A. {\bf 137} (6), pp. 696--702 (1932).

\bibitem{Jansen} S. Jansen, M. Ruskai, R. Seiler, J. Math. Phys. {\bf 48}, 102111 (2007).

\bibitem{Reichardt} B. W. Reichardt, in Proceedings of the 36th Annual ACM Symposium on Theory of Computing (STOC'04), ACM Press (2004).

\bibitem{Finnila} A. B. Finnila, M. A. Gomez, C. Sebenik, C. Stenson, J.D. Doll, Chem. Phys. Lett. {\bf 219} (1994).

\bibitem{Kadowaki} T. Kadowaki, H. Nishimori, Phys. Rev. E {\bf 58}, 5355 (1998).

\bibitem{Brady2} L. Brady, W. van Dam, Phys. Rev. A {\bf 94}, 032309 (2016).

\bibitem{Jorg} T. J\"org, F. Krzakala, J. Kurchan, A. C. Maggs, J. Pujos, Europhys. Lett. {\bf 89}, 40004 (2010).

\bibitem{Seki} Y. Seki, H. Nishimori, Phys. Rev. E {\bf 85}, 051112 (2012).

\bibitem{Bapst} V. Bapst, G. Semerjian, J. Stat. Mech. {\bf 2012}, P06007 (2012).

\bibitem{Seoane} B. Seoane, H. Nishimori, J. Phys. A {\bf 45}, 435301 (2012).

\bibitem{Susa} Y. Susa, J. F. Jadebeck, H. Nishimori, Phys. Rev. A {\bf 95}, 042321 (2017).

\bibitem{Susa2} Y. Susa, Y. Yamashiro, M. Yamamoto, H. Nishimori, J. Phys. Soc. Jpn. {\bf 87}, 023002 (2018).

\bibitem{Grover} L. K. Grover, Proc. of 28th STOC, pp. 212-219 (1996).

\bibitem{Boyer} M. Boyer, G. Brassard, P. Hoeyer, A. Tapp, Fortsch. Phys. {\bf 46}, pp. 493--506 (1998).

\bibitem{Roland} J. Roland, N. J. Cerf, Phys. Rev. A {\bf 65}, 042308 (2002).

\bibitem{vanDam02} W. van Dam, M. Mosca, U. Vazirani, Proc. of 42nd FOCS, pp. 279--287 (2001).


\end{thebibliography}
\end{document}